\documentclass[conference]{IEEEtran}
\IEEEoverridecommandlockouts
\usepackage{cite}
\usepackage{amsmath,amssymb,amsfonts}
\usepackage[inline]{enumitem}
\usepackage{booktabs}
\usepackage{colortbl}
\usepackage{graphicx}
\usepackage{textcomp}
\usepackage{csquotes}
\usepackage{xcolor}
\usepackage{subcaption}
\usepackage{microtype}
\usepackage{xurl}
\usepackage[subtle]{savetrees}
\usepackage[hidelinks]{hyperref}
\usepackage{cleveref}
\usepackage[skins,breakable]{tcolorbox}

\crefname{figure}{Fig.}{Figs.}
\Crefname{figure}{Figure}{Figures}
\crefname{section}{Sec.}{Secs.}
\Crefname{section}{Section}{Sections}
\usepackage{acronym}
\definecolor{seambg}{gray}{0.9}

\newcommand{\Seam}{\textsf{Seam}}
\newcommand{\Status}{\textsf{Status}}
\newcommand{\Trace}{\textsf{Trace}}

\acrodef{ai}[AI]{Artificial Intelligence}
\acrodef{llm}[LLM]{Large Language Model}
\acrodef{ml}[ML]{Machine Learning}

\newenvironment{infobox}[1]{
    \begin{tcolorbox}[boxrule=.5pt,
    left=1mm,
    right=1mm,
    top=1mm,
    bottom=1mm,
    ]#1
}{
    \end{tcolorbox}
}

\def\cameraready{}  
\def\BibTeX{{\rm B\kern-.05em{\sc i\kern-.025em b}\kern-.08em
    T\kern-.1667em\lower.7ex\hbox{E}\kern-.125emX}}
\begin{document}

\title{
Beyond Monolithic Models: Symbolic Seams for Composable Neuro-Symbolic Architectures

\ifdefined\cameraready
    \thanks{This work was supported by funding from the pilot program Core Informatics at KIT (KiKIT) of the Helmholtz Association (HGF) and supported by the German Research Foundation (DFG) - SFB 1608 - 501798263 and KASTEL Security Research Labs, Karlsruhe.
    }
\fi
}

\ifdefined\cameraready
    \author{
        \IEEEauthorblockN{Nicolas Schuler, Vincenzo Scotti, Raffaela Mirandola}
        \IEEEauthorblockA{\textit{KASTEL - Institute of Information Security and Dependability} \\
        \textit{Karlsruhe Institute of Technology (KIT). Germany}\\
        \{nicolas.schuler, vincenzo.scotti, raffaela.mirandola\}@kit.edu
        }
    }
\else
    \author{
        Anonymous Author(s) 
    }
\fi

\maketitle

\begin{abstract}
Current Artificial Intelligence (AI) systems are frequently built around monolithic models that entangle perception, reasoning, and decision-making, a design that often conflicts with established software architecture principles.
Large Language Models (LLMs) amplify this tendency, offering scale but limited transparency and adaptability.
To address this, we argue for composability as a guiding principle that treats AI as a living architecture rather than a fixed artifact.
We introduce \emph{symbolic seams}: explicit architectural breakpoints where a system commits to inspectable, typed boundary objects, versioned constraint bundles, and decision traces.
We describe how seams enable a composable neuro-symbolic design that combines the data-driven adaptability of learned components with the verifiability of explicit symbolic constraints -- combining strengths neither paradigm achieves alone.
By treating AI systems as assemblies of interchangeable parts rather than indivisible wholes, we outline a direction for intelligent systems that are extensible, transparent, and amenable to principled evolution.
\end{abstract}


\section{Introduction}
\label{sec:intro}

\ac{ai}\footnote{We use \emph{AI} as the umbrella term for intelligent systems. Our architectural arguments apply across the $\text{AI}\supset \text{ML} \supset \text{LLM}$ hierarchy.} has advanced through models of ever‑increasing scale, yet this progress has come at a cost to architectural discipline.
Software engineering has long emphasized modularity (principled decomposition into cohesive components)~\cite{parnas1972,baldwin1999} and composability (recombination through explicit interfaces)~\cite{dijkstra2012}.
Contemporary \ac{ai} systems, by contrast, are frequently constructed as monolithic entities that bind perception, reasoning, and decision‑making into opaque structures~\cite{card}.
In architectural terms, this design trajectory constitutes an anti‑pattern~\cite{sculley2015}: a recurring solution that undermines the principles of maintainability, transparency, and adaptability.

\acp{llm} exemplify this anti‑pattern.
As trained statistical artifacts, they demonstrate extraordinary fluency and predictive capacity, but their design reinforces a paradigm in which scale can substitute for structure and statistical power can eclipse architectural clarity.
The \textit{systems} built around such artifacts, which integrate them with data pipelines, orchestration, and control logic, inherit these structural limitations.
This creates a critical tension: the achievements of systems based on monolithic \ac{ai} components conflict with design principles that have guided software architecture for decades~\cite{lewis2024,architecture}.

The challenge, then, is not simply technical but architectural.
If \ac{ai} is to evolve beyond its current trajectory, it must be reimagined in terms of design commitments rather than sheer scale.
Symbolic \ac{ai} provides formal structure and -- to some extent -- verifiability, but struggles to learn from raw data at scale~\cite{kautz2021}. 
Data-driven approaches, e.g., \ac{ml}, deep learning, and \acp{llm}, achieve remarkable learning capacity, but sacrifice transparency and principled decomposition~\cite{garcez2023,kautz2021}.
Neuro-symbolic integration seeks to combine both strengths~\cite{belle2025}, yet the architectural foundations for composing learned and symbolic components remain underspecified.
This paper offers a first step to tackle that challenge.
Using composability as a lens, we connect established software architecture principles to neuro-symbolic designs that combine learned components with explicit symbolic artifacts (rules, constraints, logic-based reasoning) through defined interfaces.
With this first step, we outline a reference direction for systems that must evolve, repositioning \ac{ai} as a living architecture~\cite{ozkaya2022}.

This paper contributes three elements:
\begin{enumerate}[label=(\arabic*),leftmargin=*,itemsep=1pt, parsep=1pt]
    \item the concept of \emph{symbolic seams} as explicit architectural breakpoints for constraining and recomposing \ac{ai} systems,
    \item four \emph{design commitments} -- typed boundary objects, evolvable constraint configuration, externalized reasoning traces, and bounded change propagation -- that operationalize composability in neuro-symbolic settings, and
    \item a research agenda to make these commitments practical through contracts, change-impact reasoning, and constraint governance.
\end{enumerate}
To make this concrete, consider an enterprise assistant that triggers workflows under evolving constraints (access control, approvals).
In monolithic and prompt-orchestrated designs, policy updates are costly to validate: retraining may be required, or constraints remain scattered across prompts.
A symbolic seam makes boundaries explicit, enabling localized validation when policies change.

Our contribution is architectural. 
In fact, neuro-symbolic systems already employ explicit neural--symbolic interfaces, and prompt-orchestration frameworks expose intermediate steps and states~\cite{mileo2025,badreddine2022a,langchain,dspy}.
Conversely, we specify seams as explicit connectors, shifting composability from \enquote{modular execution} to~\enquote{governed recombination under stable contracts}, regardless of whether components are neural, symbolic, or hybrid.

\section{Background and Related Work}

Our argument draws on four research areas: \begin{enumerate*}[label=(\arabic*)]
  \item foundational architectural principles that make systems evolvable, 
  \item empirical evidence that \ac{ml} systems violate those principles,
  \item prompt-orchestration frameworks that modularize execution without governing interaction semantics, and 
  \item neuro-symbolic approaches that restore composable structure but leave a connector-level gap.
\end{enumerate*}

\textit{Architectural foundations.}
Software engineering rests on foundational principles established through decades of empirical research.
Examples include Parnas's information hiding~\cite{parnas1972}, Dijkstra's separation of concerns~\cite{dijkstra2012}, and Baldwin and Clark's theory of modularity~\cite{baldwin1999}, collectively demonstrating that principled decomposition improves flexibility and comprehensibility.
Maintainability, transparency, and adaptability emerge not coincidentally, but systematically from such decomposition.

\textit{Technical debt in \ac{ml} systems.}
\ac{ml} systems incur unique forms of technical debt~\cite{sculley2015}, including what the authors term \textit{abstraction debt}: the difficulty of enforcing strict abstraction boundaries, which degrades modularity.
Empirical studies confirm these risks: \ac{ml} repositories exhibit higher levels of technical debt than non-\ac{ml} projects~\cite{alfadel2023}, often characterized by weak component boundaries and fragile orchestration (\enquote{glue code})~\cite{nazir2024}.
While recent research proposes methods to detect \ac{ml}-specific code smells~\cite{recupito2024}, these approaches often treat symptoms of the underlying architectural tension rather than resolving it.

\textit{Prompt-orchestration frameworks.}
Frameworks such as LangChain~\cite{langchain} and DSPy~\cite{dspy} respond pragmatically to this tension: they decompose \ac{ai} application logic into pipelines of discrete steps (prompting, retrieval, tool invocation), thereby modularizing flow and exposing intermediate states for inspection.
Yet, the constraints governing interaction between steps (input and output schemas, validation rules, consistency requirements) typically remain embedded in prompt templates and ad-hoc checks rather than in governed interfaces.

\textit{Neuro-symbolic approaches and composability.}
Recent work on neuro-symbolic \ac{ai} emphasizes explicit interfaces between neural and symbolic components, motivated in part by concerns that end-to-end integration conflicts with architectural principles that have long proven effective in software engineering and symbolic \ac{ai}~\cite{mileo2025}.
Neuro-symbolic systems typically partition functionality into distinct layers with defined interfaces: neural components extract patterns from unstructured data, while symbolic components apply logical inference and rule-based constraints to those abstractions~\cite{belle2025,badreddine2022a}.
Researchers emphasize that this separation exposes human-readable reasoning traces and supports component-level replacement and evolution~\cite{mileo2025}.
By contrast, purely neural modular designs can also be composed, but their interfaces are often statistical and may require additional alignment when modules are replaced.
Existing frameworks such as Logic Tensor Networks~\cite{badreddine2022a} and DeepProbLog~\cite{robin2018} demonstrate that composable neuro-symbolic designs are implementable and can achieve strong performance where interpretability or domain constraints matter.

\textit{Related work and gap.}
Across these four lines of work, the literature identifies symptoms and provides building blocks, but it does not close the architectural loop: it lacks a connector-level specification that turns inevitable change into bounded, checkable revalidation. Work on \ac{ml} technical debt, entanglement diagnoses, and glue-code fragility exists~\cite{sculley2015,alfadel2023,nazir2024}, yet offers no notion of \emph{seam-level contracts} that scope the impact of updates. Neuro-symbolic frameworks introduce explicit interfaces for injecting constraints~\cite{mileo2025,belle2025,badreddine2022a}, but typically treat constraints and traces as task-bound mechanisms rather than \emph{versioned, governed artifacts} that persist across system evolution. Prompt-orchestration frameworks, despite modularizing execution, leave interaction semantics informal and therefore difficult to validate or evolve systematically.
Self-adaptive systems (SAS) frameworks such as MAPE-K~\cite{kephart2003} formalize adaptation loops with architectural models, and ADLs such as Wright~\cite{allen1997} specify typed connector protocols. These provide general-purpose adaptation mechanisms but target managed components with predictable behavior; they do not address the stochastic outputs, distributional contracts, and retraining-driven evolution specific to learned components.
Symbolic seams close this gap by elevating boundary objects, constraint bundles, and decision traces into \emph{durable connector specifications}: changes are mediated through seam versions, triggering localized validation rather than system-wide revalidation. 

\section{Why Monolithic AI Resists Architectural Evolution}
\begin{figure*}[t!]
  \centering

  \begin{subfigure}[b]{0.45\textwidth}
    \centering
    \includegraphics[width=\linewidth]{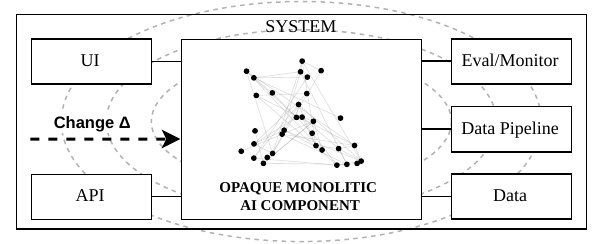}
    \caption{Monolitic \ac{ai} system}
    \label{fig:panelA}
  \end{subfigure}
  \hfill
  \begin{subfigure}[b]{0.45\textwidth}
    \centering
    \includegraphics[width=\linewidth]{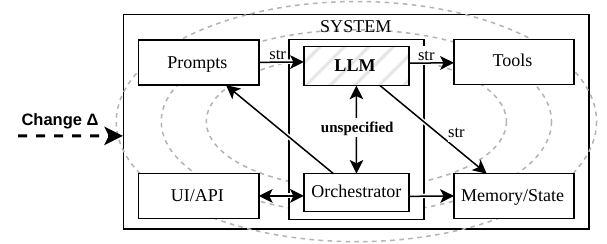}
    \caption{Prompt-Orchestrated System}
    \label{fig:panelB}
  \end{subfigure}

  \caption{Current entanglement postures: (a)~system is substantially overlapped with its neural logic; (b)~entanglement is hidden through informal mechanisms. $\Delta$s represent the changes that affect the system.}
  \label{fig:overview}
\end{figure*}
Before introducing symbolic seams, we examine what makes monolithic architectures brittle.
The core pattern is clear; \textit{Symptom}: small changes require broad revalidation; \textit{Cause}: latent coupling from concerns fused in a single parametric space; \textit{Consequence}: expensive evolution and brittle governance.
Sculley et al. first articulated the \textit{Changing Anything Changes Everything} (CACE) phenomenon in their study of \ac{ml} technical debt~\cite{sculley2015}.
While their core argument was framed as the accumulation of technical debt, CACE also points to a deeper architectural problem: \textit{entanglement} in end-to-end trained \ac{ml} models.
Entanglement means that multiple concerns are encoded jointly, so changing one aspect perturbs others in hard-to-predict ways, undermining \textit{architectural scalability} (i.e., the ability to absorb change with bounded impact and diagnosable consequences).

Consider an end-to-end trained \ac{ml} component such as an \ac{llm}: representations, intermediate computations, and output selection are optimized together against a single global loss function.
Changes to learned representations (e.g., modifying how entities are encoded) can cascade through downstream computations, which in turn alter decision-making behavior. 
This coupling is structural: upstream changes shift downstream input distributions, potentially invalidating learned mappings throughout the system~\cite{sculley2015}.
\Cref{fig:panelA} exemplifies this posture: perception, reasoning, and decision-making are collapsed into a single statistical artifact and its surrounding \enquote{glue} -- an extreme case of entanglement.

This structural entanglement conflicts with at least three foundational software architecture principles.
\textit{Separation of Concerns} becomes difficult to enforce when concerns are fused in a single parametric space: responsibilities cannot be cleanly isolated because they share optimization objectives and parameter updates.
\textit{Information Hiding} is weakened because the decision pathway remains opaque -- there is often no clean boundary where intermediate reasoning is exposed for inspection or validation~\cite{rudin2021,rudin2019}.
\textit{Composability} becomes difficult to achieve: upgrading or swapping one component (e.g., perception) can cascade through reasoning and decision logic, locking the system in mutual dependence.
These problems are not merely implementation details to be fixed through better engineering, but constraints imposed by how monolithic \ac{ai} systems are built and optimized end-to-end~\cite{bhatia2024}. 

This rigidity manifests in three critical deficiencies:
\begin{enumerate}[label=(\arabic*),leftmargin=*]
    \item \textit{Explainability}: decisions emerge from opaque parameter spaces; post-hoc methods typically offer local approximations rather than faithful explanations and remain limited under entanglement~\cite{rudin2019,rudin2021}.
    \item \textit{Adaptability}: model and concept drift (degradation as real-world data distributions shift) often drive expensive full retraining rather than localized component-level updates~\cite{vela2022,glasmachers2017}.
    \item \textit{Composability}: perception, reasoning, and decision components are difficult to swap independently, forcing monolithic development or ad-hoc composition without principled interfaces~\cite{pan2020}.
\end{enumerate}
Prompt-orchestration frameworks (\Cref{fig:panelB}) attempt to address these deficiencies by modularizing execution flows. 
Yet, the critical semantics of intermediate states and constraints often remain informal and entangled in prompts and ad-hoc checks.

This raises an architectural question: can we reimagine \ac{ai} systems as decomposable structures in which perception, reasoning, and decision-making are articulated as interoperable, independently evolvable layers?

\section{Symbolic Seams for Composable Architecture}
\begin{figure*}[!ht]
    \centering
    \begin{subfigure}[]{0.45\textwidth}
        \centering
        \includegraphics[width=\linewidth]{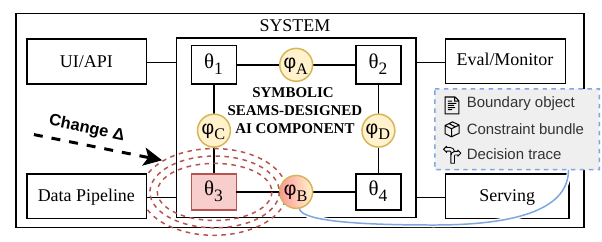}
        \caption{Symbolic Seams Posture}
        \label{fig:panelC}
    \end{subfigure}
    \hfill
    \begin{subfigure}[]{0.45\textwidth}
        \centering
        \includegraphics[width=\linewidth]{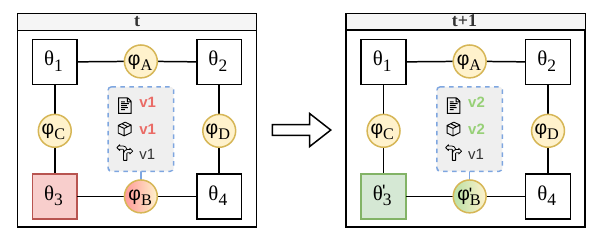}
        \caption{Change Propagation}
        \label{fig:evolution}
    \end{subfigure}
    \hfill
    \caption{Proposed posture: (a)~learned modules ($\theta$) are separated by symbolic seam connectors ($\phi$) that checkpoint constraints; the dashed circle shows change bounded at the affected seam. (b)~Seams persist across evolution while modules are replaced ($\theta'$ and $\phi'$ are the updated versions). Effect of change $\Delta$ is contained.}
    \label{fig:seam}
\end{figure*}

Decomposition is the natural response to architectural brittleness introduced by entanglement, but it requires stable boundaries that preserve the ability to reason about, evolve, and govern a system under change.
As \cref{fig:overview} showed, neither the monolithic nor the prompt-orchestrated posture provides such boundaries.

\subsection{Overview}
\label{subsec:overview}

\begin{figure}[!ht]
\centering
\setlength{\abovedisplayskip}{2pt}
\setlength{\belowdisplayskip}{2pt}
\begin{equation*}
\label{eq:seam-contract}
\begin{aligned}
\Seam(\phi) \;&\triangleq\;
  \big\langle
    \mathcal{T}_{\mathrm{in}},\,
    \mathcal{T}_{\mathrm{out}},\,
    \mathcal{C}^{(v)},\,
    \mathcal{S}_{\mathrm{trace}}
  \big\rangle \\
\phi:\; \mathcal{T}_{\mathrm{in}} \times \mathcal{C}^{(v)}
\;&\longrightarrow\;
  \mathcal{T}_{\mathrm{out}} \times \Status \times \Trace \\
\Trace \;&\triangleq\;
  \big\langle v,\, \textsf{evidence} \big\rangle
\end{aligned}
\end{equation*}
\caption{A seam ($\phi$) commits to typed boundary objects (\(\mathcal{T}_{\mathrm{in}}, \mathcal{T}_{\mathrm{out}}\)), a versioned constraint bundle \(\mathcal{C}^{(v)}\), and a trace schema \(\mathcal{S}_{\mathrm{trace}}\). For stochastic components, the contract may specify distributional properties.}
\label{fig:seam-contract}
\end{figure}

\textit{Symbolic seams} are our envisoned solution to tackle the challenges of current \ac{ai}-based systems. 
\begin{infobox}
A symbolic seam is a connector-level constraint checkpoint that emits typed boundary objects, evaluates versioned constraint bundles, and records decision traces.
\end{infobox}
Seams are not components or modules themselves, but commitments that externalize constraints and trace decisions \emph{between} components.
\Cref{fig:panelC} makes this architecture concrete.
Learned modules ($\theta_1$--$\theta_4$) handle data-driven tasks, perception, classification, generation, while symbolic seam connectors ($\phi_A$--$\phi_D$) mediate every crossing between them.
Each~$\phi$ validates boundary objects against the active constraint bundle and emits a decision trace before data passes to the next module.
When a change arrives (change~$\Delta$ in the figure), its impact is bounded by the surrounding seams: the dashed circle around $\theta_3$ shows that only the affected module and its adjacent contracts require revalidation, leaving the rest of the assembly unaffected.

\Cref{fig:evolution} illustrates the temporal consequence.
On the left, modules~$\theta_1$--$\theta_4$ are wired through seams~$\phi_A$--$\phi_D$.
On the right, after a change occurred,~$\theta_3$ has been replaced by an updated version (marked~$\theta_3^{'}$, denoting a retrained, fine-tuned, or swapped variant) with a versioned connector level update of~$\phi_B^{'}$.
Seams form the system's durable, versioned skeleton, providing stable contract behavior against which each new module version can be validated independently.
Seams typically wrap learnable components, but constraints can also be partially internalized when they shape a component's structure or learning objective.
They behave as connector contracts (\Cref{fig:seam-contract}) -- explicit interfaces that bundle typed boundary objects, constraint configuration, and trace schemas to govern interaction and change propagation. They guarantee checkability of the declared interface but not completeness of all downstream reasoning.
\Cref{fig:seam-contract} scopes the seam's interface commitments, not a full formal semantics. For stochastic components, ``typed'' means that boundary objects declare not only structure (fields, types, ranges) but also distributional properties (e.g., calibration bounds, coverage guarantees) that the seam can validate at the interface.

\begin{table}[!ht]
\setlength{\aboverulesep}{0pt}
\setlength{\belowrulesep}{0pt}
\setlength{\tabcolsep}{4pt}
  \centering
  \footnotesize
  \begin{tabular}{lp{1.8cm}p{1.7cm}>{\columncolor{seambg}}p{2cm}}
    \toprule
    & \textbf{Model-Centric} & \textbf{Prompt-Orch.} & \textbf{Sym. Seams} \\
    \midrule
    Contracts & Implicit          & Informal (str) & Typed, versioned \\
    Evolution & Retrain all       & Edit prompts   & Seam-level config \\
    Change visibility & None      & Unclear        & Seam validation   \\
    \bottomrule
  \end{tabular}
  \caption{Comparison of entanglement postures.}
  \label{tab:entanglement-comparison}
\end{table}

Symbolic seams differ from prior work in three respects, summarized in \Cref{tab:entanglement-comparison}:
\begin{enumerate*}[label=(\arabic*)]
    \item they specify interaction at the \emph{connector} level rather than as component-level mechanisms,
    \item they treat constraints and traces as \emph{versioned, governed artifacts} rather than task-specific logic,
    \item they make \emph{bounded change propagation} an explicit architectural property rather than an emergent side effect.
\end{enumerate*}
The shift is from exchanging intermediate representations to governing interaction, evolution, and validation at architectural boundaries. 

\subsection{Design Commitments}
\label{sec:commitments}

We propose four design commitments that specify the interface properties a symbolic seam must uphold.

\textbf{Typed, Inspectable Boundary Objects:} Components communicate via explicit boundary objects with declared structure and semantics, rather than through implicit latent couplings.
A boundary object may be symbolic, neural, or hybrid, but it must be inspectable and testable as an interface artifact~\cite{card,architecture}.

\textbf{Evolvable Constraint Configuration:} Behavioral constraints (policies, domain rules, safety boundaries) are externalized as first-class configuration artifacts that can be injected, modified, and rolled back without full end-to-end retraining.
This turns many behavior changes from a re-optimization problem into a governed configuration change.

\textbf{Externalized Reasoning Traces (Decision Receipts):} The system emits intermediate reasoning checkpoints during normal operation, producing bounded, checkable traces for debugging, audit, and governance.
Traces are structured execution artifacts (e.g., constraint checks and evidence references), not necessarily free-form natural-language rationales.
This makes transparency a system property rather than a retroactive interpretation of opaque outputs.

\textbf{Bounded Change Propagation:} Symbolic seams function as abstraction barriers, where changes behind a seam have diagnosable and bounded impact on other components unless the interface changes.
This enables localized evolution (e.g., replacing a perception component) while preserving system-level predictability.

\subsection{Illustrative Scenario: Enterprise Policy Assistant}
\label{sec:scenario}

Considering the enterprise assistant sketched in \Cref{sec:intro} that answers questions, drafts tickets, and triggers workflows under evolving constraints, in a model-centric design, constraints are absorbed into training data, fine-tuning, or monolithic prompting patterns, making updates costly and difficult to validate. 
In an orchestrated design, policies are implemented as prompt fragments and scattered checks; workflows are modularized, but constraint semantics remain fragile and difficult to govern.

Under symbolic seam commitments, the assistant is organized around explicit contracts at connector boundaries: each stage emits a typed boundary object that can be validated and logged, rather than keeping semantics implicit in prompts.
Boundary objects and traces live \emph{at the seam}: components may produce internal intermediate representations, but the seam commits to externalized, inspectable artifacts. It governs validation, versioning, and trace emission, even when hybrid components emit boundary objects directly.
When a policy changes (e.g., a new approval rule), the update becomes a constraint-bundle version change with seam-level regression checks over boundary objects and decision records.
If the boundary-object contract remains stable, learned components can be swapped or retrained behind the seam without rewriting downstream policy logic.
For a policy update, the postures diverge sharply: monolithic approaches require retraining with full regression; in prompt-orchestrated approaches, edits affect prompts without validation contracts; symbolic seams-based approaches version a constraint bundle with seam-scoped regression and rollback.

Symbolic seams primarily make run-time constraints explicit and governable, though the same commitments support compile-time approaches where constraints shape a component's learning objective~\cite{logicalrbm,comp}.
Seen through a composability lens, seams enable three disciplined operations:
\begin{enumerate*}[label=(\arabic*)]
    \item \textit{evolve} behavior by editing and versioning constraints,
    \item \textit{replace} learnable components behind a seam as long as their boundary-object contract remains stable, and
    \item \textit{recompose} workflows by wiring components through typed boundary objects rather than implicit prompt orchestration.
\end{enumerate*}
This highlights a key distinction: modularity decomposes execution, whereas composability enables safe recombination under explicit contracts.
Components can be replaced, reordered, or constrained with bounded validation as long as the seam contract remains stable.

Existing neuro-symbolic approaches demonstrate that neural learning can be combined with explicit symbolic interfaces~\cite{badreddine2022a,robin2018}, yet these approaches remain largely domain-specific and require substantial expertise to deploy at scale.
The commitments above are architectural standards rather than recipes: closing the gap between these principles and practical tooling remains an open challenge.

\section{Discussion and Research Agenda}

Symbolic seams offer a complementary direction to the current emphasis on scale and end-to-end optimization,  treating constraints, interfaces, and traces as durable architectural artifacts rather than addressing every change through retraining or prompt engineering.
Unlike post-hoc explainability methods that infer rationales after the fact, seam-level traces are execution artifacts: intermediate decisions and constraint evaluations are emitted during normal operation and can be validated at the interface, making trace fidelity an architectural requirement.
This shifts the question from~\enquote{why did the model say this?} to~\enquote{did the system follow the specified constraints and contracts?}, turning verification and validation into the primary assurance mechanism. 
Rather than relying on interpretability alone, engineers can check contract satisfaction at each seam, validate trace completeness against declared decision points, and regression-test constraint enforcement when components change.

To make the above commitments actionable, we outline a research agenda focused on turning seams, constraints, and traces into engineering artifacts with the following concrete challenges:
\begin{enumerate}[label=(\arabic*),leftmargin=*,itemsep=1pt, parsep=1pt]
    \item Interface formalization: How can contracts capture stochastic outputs while remaining checkable?
    \item Change-impact reasoning: How can engineers estimate the behavioral impact of changing constraints or replacing components without exhaustive testing or expensive retraining?
    \item Operational observability and trace fidelity: How can we ensure reasoning traces remain faithful to the actual decision process and operationally reliable as models, constraints, and components evolve?
    \item Tooling for constraint governance: What are the analogues of version control, diffing, and rollback for constraint specifications and their enforced implementations?
\end{enumerate}
Evaluation can borrow from self-adaptive systems by treating policy updates and component swaps as adaptation scenarios and measuring change locality, assurance effort, and contract breakage rate across architectural postures.

That said, composability does not come for free: end-to-end optimization can yield higher raw performance precisely because it exploits cross-component entanglement as a learnable parameter.
The architectural question is therefore not whether to decompose, but where to place seams so that the benefits of adaptability, explainability, and governance outweigh the cost of explicit interfaces and constraints -- a central trade-off of treating \ac{ai} as a living architecture.

\bibliographystyle{IEEEtran}
\bibliography{IEEEabrv,bib}

@article{sculley2015,
  title={Hidden technical debt in machine learning systems},
  author={Sculley, David and others},
  journal={NeurIPS},
  volume={28},
  year={2015}
}

@article{rudin2021,
  author       = {Cynthia Rudin and
                 others},
  title        = {Interpretable Machine Learning: Fundamental Principles and 10 Grand
                  Challenges},
  journal      = {CoRR},
  volume       = {abs/2103.11251},
  year         = {2021},
  eprinttype    = {arXiv},
  eprint       = {2103.11251},
}

@article{vela2022,
author={Vela, Daniel
and Sharp, Andrew
and Zhang, Richard
and Nguyen, Trang
and Hoang, An
and Pianykh, Oleg S.},
title={Temporal quality degradation in AI models},
journal={Scientific Reports},
year={2022},
month={Jul},
day={08},
volume={12},
number={1},
pages={11654},
issn={2045-2322},
doi={10.1038/s41598-022-15245-z},
}

@article{badreddine2022a,
title = {Logic Tensor Networks},
journal = {Artificial Intelligence},
volume = {303},
pages = {103649},
year = {2022},
issn = {0004-3702},
doi = {https://doi.org/10.1016/j.artint.2021.103649},
author = {Samy Badreddine and others}
}

@online{bhatia2024,
  title = {An {{Empirical Study}} of {{Self-Admitted Technical Debt}} in {{Machine Learning Software}}},
  author = {Bhatia, Aaditya and others},
  date = {2024-06-09},
  eprint = {2311.12019},
  eprinttype = {arXiv},
  eprintclass = {cs},
  doi = {10.48550/arXiv.2311.12019},
  langid = {english},
  pubstate = {prepublished},
}

@article{rudin2019,
author={Rudin, Cynthia},
title={Stop explaining black box machine learning models for high stakes decisions and use interpretable models instead},
journal={Nature Machine Intelligence},
year={2019},
month={May},
day={01},
volume={1},
number={5},
pages={206-215},
issn={2522-5839},
doi={10.1038/s42256-019-0048-x},
}

@Article{lewis2024,
  author =	{Lewis, Grace A. and others},
  title =	{{Software Architecture and Machine Learning (Dagstuhl Seminar 23302)}},
  pages =	{166--188},
  journal =	{Dagstuhl Reports},
  ISSN =	{2192-5283},
  year =	{2024},
  volume =	{13},
  number =	{7},
  publisher =	{Schloss Dagstuhl -- Leibniz-Zentrum f{\"u}r Informatik},
  address =	{Dagstuhl, Germany},
  URN =		{urn:nbn:de:0030-drops-197793},
  doi =		{10.4230/DagRep.13.7.166}
}

@InProceedings{glasmachers2017,
  title = 	 {Limits of End-to-End Learning},
  author = 	 {Glasmachers, Tobias},
  booktitle = 	 {Proceedings of the Ninth Asian Conference on Machine Learning},
  pages = 	 {17--32},
  year = 	 {2017},
  volume = 	 {77},
  address = 	 {Yonsei University, Seoul, Republic of Korea},
  month = 	 {15--17 Nov},
  publisher =    {PMLR}
}

@inproceedings{pan2020,
  title = {On Decomposing a Deep Neural Network into Modules},
  booktitle = {FSE},
  author = {Pan, Rangeet and Rajan, Hridesh},
  date = {2020-11-08},
  pages = {889--900},
  publisher = {ACM},
  year = {2020},
  location = {Virtual Event USA},
  doi = {10.1145/3368089.3409668},
  isbn = {978-1-4503-7043-1},
  langid = {english}
}

@article{parnas1972,
author = {Parnas, D. L.},
title = {On the criteria to be used in decomposing systems into modules},
year = {1972},
issue_date = {Dec. 1972},
publisher = {Association for Computing Machinery},
address = {New York, NY, USA},
volume = {15},
number = {12},
issn = {0001-0782},
doi = {10.1145/361598.361623},
journal = {Commun. ACM},
month = dec,
pages = {1053–1058},
numpages = {6}
}

@book{dijkstra2012,
  title={Selected writings on computing: a personal perspective},
  author={Dijkstra, Edsger W},
  year={2012},
  publisher={Springer Science \& Business Media}
}

@book{baldwin1999,
author = {Baldwin, Carliss Y. and Clark, Kim B.},
title = {Design Rules: The Power of Modularity Volume 1},
year = {1999},
isbn = {0262024667},
publisher = {MIT Press},
address = {Cambridge, MA, USA}
}

@inproceedings{alfadel2023,
author = {OBrien, David and others},
title = {23 shades of self-admitted technical debt: an empirical study on machine learning software},
year = {2022},
isbn = {9781450394130},
publisher = {Association for Computing Machinery},
address = {New York, NY, USA},
doi = {10.1145/3540250.3549088},
booktitle = {FSE},
pages = {734–746},
numpages = {13},
location = {Singapore, Singapore},
series = {ESEC/FSE 2022}
}

@article{nazir2024,
title = {Architecting ML-enabled systems: Challenges, best practices, and design decisions},
journal = {Journal of Systems and Software},
volume = {207},
year = {2024},
issn = {0164-1212},
doi = {https://doi.org/10.1016/j.jss.2023.111860},
author = {Roger Nazir and others}
}

@article{recupito2024,
title = {Technical debt in AI-enabled systems: On the prevalence, severity, impact, and management strategies for code and architecture},
journal = {Journal of Systems and Software},
volume = {216},
year = {2024},
issn = {0164-1212},
doi = {https://doi.org/10.1016/j.jss.2024.112151},
author = {Gilberto Recupito and others}
}

@article{mileo2025,
author = {Alessandra Mileo},
title ={Towards a neuro-symbolic cycle for human-centered explainability},

journal = {Neurosymbolic Artificial Intelligence},
volume = {1},
number = {},
pages = {NAI-240740},
year = {2025},
doi = {10.3233/NAI-240740},
eprint = { 
    
        https://doi.org/10.3233/NAI-240740
    
    

}
}

@article{garcez2023,
  title = {Neurosymbolic {{AI}}: The 3rd Wave},
  author = {Garcez, Artur d'Avila and Lamb, Lu{\'i}s C.},
  year = {2023},
  journal = {Artificial Intelligence Review},
  volume = {56},
  number = {11},
  pages = {12387--12406},
  issn = {1573-7462},
  doi = {10.1007/s10462-023-10448-w}
}

@article{kautz2021,
author = {Kautz, Henry A.},
title = {The third AI summer: AAAI Robert S. Engelmore Memorial Lecture},
journal = {AI Magazine},
volume = {43},
number = {1},
pages = {105-125},
doi = {https://doi.org/10.1002/aaai.12036},
eprint = {https://onlinelibrary.wiley.com/doi/pdf/10.1002/aaai.12036},
year = {2022}
}

@article{belle2025,
author = {Vaishak Belle},
title ={On the Relevance of Logic for Artificial Intelligence, and the Promise of Neurosymbolic Learning},

journal = {Neurosymbolic Artificial Intelligence},
volume = {1},
number = {},
year = {2025},
doi = {10.1177/29498732251339951},
eprint = { 
    
        https://doi.org/10.1177/29498732251339951
    
    

}
}

@inproceedings{robin2018,
 author = {Manhaeve, Robin and others},
 booktitle = {NeurIPS},
 pages = {},
 publisher = {Curran Associates, Inc.},
 title = {DeepProbLog:  Neural Probabilistic Logic Programming},
 volume = {31},
 year = {2018}
}

@ARTICLE{ozkaya2022,
  author={Ozkaya, Ipek},
  journal={IEEE Software}, 
  title={An AI Engineer Versus a Software Engineer}, 
  year={2022},
  volume={39},
  number={6},
  pages={4-7},
  doi={10.1109/MS.2022.3161756},
  ISSN={1937-4194},
  month={Nov},
}

@article{dspy,
  author       = {Omar Khattab and others},
  title        = {DSPy: Compiling Declarative Language Model Calls into Self-Improving
                  Pipelines},
  journal      = {CoRR},
  volume       = {abs/2310.03714},
  year         = {2023},
  doi          = {10.48550/ARXIV.2310.03714},
  eprinttype    = {arXiv},
  eprint       = {2310.03714},
}

@misc{langchain,
author = {Chase, Harrison},
month = oct,
title = {{LangChain}},
url = {https://github.com/langchain-ai/langchain},
year = {2022}
}

@article{logicalrbm,
  author       = {Son N. Tran and
                  Artur S. d'Avila Garcez},
  title        = {Logical Boltzmann Machines},
  journal      = {CoRR},
  volume       = {abs/2112.05841},
  year         = {2021},
  eprinttype    = {arXiv},
  eprint       = {2112.05841},
}

@inproceedings{card,
author = {Mitchell, Margaret and others},
title = {Model Cards for Model Reporting},
year = {2019},
isbn = {9781450361255},
publisher = {Association for Computing Machinery},
address = {New York, NY, USA},
doi = {10.1145/3287560.3287596},
booktitle = {Proceedings of the Conference on Fairness, Accountability, and Transparency},
pages = {220–229},
numpages = {10},
location = {Atlanta, GA, USA},
series = {FAT* '19}
}

@inproceedings{comp,
  author       = {Yilun Du and
                  others},
  title        = {Compositional Visual Generation with Energy Based Models},
  booktitle    = {NeurIPS 2020, December
                  6-12, 2020, virtual},
  year         = {2020},
}

@book{architecture,
  author       = {Leonard J. Bass and
                  others},
  title        = {Software architecture in practice},
  series       = {{SEI} series in software engineering},
  publisher    = {Addison-Wesley-Longman},
  year         = {1999},
  isbn         = {978-0-201-19930-7},
}

@ARTICLE{kephart2003,
  author={Kephart, J.O. and Chess, D.M.},
  journal={Computer}, 
  title={The vision of autonomic computing}, 
  year={2003},
  volume={36},
  number={1},
  pages={41-50},
  doi={10.1109/MC.2003.1160055}}

@article{allen1997,
author = {Allen, Robert and Garlan, David},
title = {A formal basis for architectural connection},
year = {1997},
issue_date = {July 1997},
publisher = {Association for Computing Machinery},
address = {New York, NY, USA},
volume = {6},
number = {3},
issn = {1049-331X},
doi = {10.1145/258077.258078},
journal = {ACM Trans. Softw. Eng. Methodol.},
month = jul,
pages = {213–249},
numpages = {37}
}
\end{document}